\def\ra{\rangle}

\documentclass[aps,amsmath,amssymb,amsfonts,showpace,twocolumn,pra]{revtex4}
\usepackage{amscd}
\usepackage{amsthm}
\usepackage{graphicx}
\usepackage[
  colorlinks,
  linkcolor = blue,
  citecolor = blue,
  urlcolor = blue]{hyperref}

\newcommand{\tabincell}[2]{\begin{tabular}{@{}#1@{}}#2\end{tabular}}

\def\qed{\leavevmode\unskip\penalty9999 \hbox{}\nobreak\hfill
     \quad\hbox{\leavevmode  \hbox to.77778em{%
               \hfil\vrule   \vbox to.675em%
               {\hrule width.6em\vfil\hrule}\vrule\hfil}}
     \par\vskip3pt}

\newtheorem{theorem}{Theorem}
\newtheorem{corollary}{Corollary}
\newtheorem{lemma}{Lemma}

\usepackage{epstopdf}
\usepackage{pifont}
\usepackage{epsfig,subfigure,dsfont,amsthm,amsbsy,mathrsfs,amscd}
\usepackage{caption}
\def\qed{\hfill \vrule height7pt width 7pt depth 0pt}
\begin{document}
\title{Constructing   unextendible product bases from the old ones}
\author{Yan-Ling Wang$^1$, Mao-Sheng Li$^2$, Shao-Ming Fei$^{3,4}$, Zhu-Jun Zheng$^1$}

\affiliation{
$^1$School of Mathematics, South China University of Technology, Guangzhou 510640, China\\
$^2$School of Mathematical Sciences,
 Tsinghua University, Beijing
100084, China\\
$^3$School of Mathematical Sciences, Capital Normal
University, Beijing 100048, China\\
$^4$Max-Planck-Institute for Mathematics in the
Sciences, 04103 Leipzig, Germany}

\begin{abstract}
We studied the construction problem of the unextendible product basis (UPB). We mainly give a method to construct  a UPB of a quantum system through the UPBs of its subsystem. Using this method and the UPBs which are known for us, we construct different kinds of  UPBs in general bipartite quantum system. Then we use these UPBs to construct a family of UPBs in multipartite quantum system. The UPBs can be used to construct the bound entangled states with different ranks.

\end{abstract}

\maketitle

\section{Introduction}

The notion of unextendible product bases (UPBs) plays an important role in quantum information theory. Since it was
been proposed in 1999 by Bennett \emph{et al}\cite{BDM+99}, it   has been extensively investigated. Considerable elegant results have been
obtained with interesting applications to the theory of quantum information. Such as the construction of bound entangled states\cite{BDM+99,LSM11,S11} and indecomposible positive maps\cite{T01}. And UPB have also been shown to give rise to Bell inequalities without a quantum violation\cite{ASH+11} and they can not be perfectly distinguished by local quantum operations and classical communication, even though they contain no entanglement\cite{BDF+99}. The UPB is a set of incomplete
orthonormal product states whose complementary space has no product states.
It was also shown that the space complementary to a UPB contains bound entanglement \cite{H97}.


In the past years, there also have many results about the construction of UPBs. The first UPB is a set with five product states in $\mathbb{C}^{3}\bigotimes\mathbb{C}^{3}$ which was constructed by Bennett \emph{et al} in 1999 \cite{BDM+99}. Then in \cite{DMS+03}, the authors used the orthogonal graph to construct a UPB in $\mathbb{C}^m\otimes\mathbb{C}^n(m\geq3, n>3, m\leq n)$ with number $mn-2m+1$. It is interesting to characterize the minimal number of product states in order to form a UPB. In \cite {AL01}, the authors showed that in $\mathbb{C}^{d_1}\otimes\mathbb{C}^{d_2}\ldots\otimes\mathbb{C}^{d_m}$ the minimum size of UPB is $\sum_{i=1}^m(d_i-1)+1$ for every sequence of integers $d_1,d_2,\ldots,d_m\geq 2$ unless either (i) $m=2$ and $2\in \{d_1,d_2\}$ or (ii) $\sum_{i=1}^m(d_i-1)+1$ is odd and at least one $d_i$ is even. In 2015, the authors further determined the unsolved minimum number  in bipartite cases as well as a large family of multipartite cases\cite{CJ15}. Moreover, Johnston focused on the structure of qubit UPB and obtained a class of UPB with different number of states\cite{J14}. However, there are only a few cases whose structure of UPBs were completely characterized: $\mathbb{C}^{2}\bigotimes\mathbb{C}^{n},$ $\mathbb{C}^{3}\bigotimes\mathbb{C}^{3},$
$\mathbb{C}^{2}\bigotimes\mathbb{C}^{2}\bigotimes\mathbb{C}^2,$ and $\mathbb{C}^{2}\bigotimes\mathbb{C}^{2}\bigotimes\mathbb{C}^2\bigotimes\mathbb{C}^2$ \cite{BDM+99,DMS+03,B04,J14}. More recently, the authors focused on the characterization of the UPB in $\mathbb{C}^3\bigotimes\mathbb{C}^4$ \cite{YGX+15}.

From \cite{J14,DMS+03}, the author derived some methods to obtain a UPB of higher dimensional system via some small systems. For example, it was showed that the tensor product of UPBs is again a UPB of higher dimensional system. That is, given two bipartite UPBs $S_1=\{|\psi_i^1\rangle, i=1,2,\ldots,l_1\}$ in $\mathbb{C}^{n_1}\otimes\mathbb{C}^{m_1}$ and $S_2=\{|\psi_i^2\rangle, i=1,2,\ldots,l_2\}$ in $\mathbb{C}^{n_2}\otimes\mathbb{C}^{m_2}$, the states $\{|\psi_i^1\rangle\otimes|\psi_j^2\rangle\}_{i,j=1}^{l_1,l_2}$ are again product states which is proved to form   a bipartite UPB on $\mathbb{C}^{n_1n_2}\otimes\mathbb{C}^{m_1m_2}$. Hence in order to characterize the UPBs, it is worth to give more methods to obtain some new UPBs.

In this paper, we mainly pay attention to the constructions of  UPBs in bipartite system. Embedding the bipartite systems $\mathbb{C}^{m}\bigotimes\mathbb{C}^{n_1}$ and $\mathbb{C}^{m}\bigotimes\mathbb{C}^{n_2}$  as two orthogonal subsystems of $\mathbb{C}^{m}\bigotimes\mathbb{C}^{n_1+n_2}$, we  proved that the union of UPBs of the two subsystems is again a UPB of the whole system. Then we used this method to construct lots of UPBs in general  bipartite quantum system $\mathbb{C}^{m}\bigotimes\mathbb{C}^{n}$ from some haven knew results. More precisely,   we used this method to construct  a large number of UPBs in $\mathbb{C}^{n}\bigotimes\mathbb{C}^{n}$ whose missing states number of states are varying from $4$ to $\frac{n^2}{2}$. (Here we specific explain the words "missing states number": If the set $S=\{|\psi_i\rangle\mid i=1,2,\ldots,k, k<mn\}$ is a UPB in $\mathbb{C}^{m}\bigotimes\mathbb{C}^{n}$, then $mn-k$ is the  missing states number of $S$, it characterize how far is the UPBs in order to form a base of the whole system.)
Moreover, for the general bipartite quantum system $\mathbb{C}^{m}\bigotimes\mathbb{C}^{n}(m\geq4,  n \geq10)$, we obtained the UPBs with missing number varying  from $4$ to $(m-1)(n-8).$ At last, we also pay our attention to the multipartite quantum systems and give a method to derive a UPB from UPB in less partite system. Then we use the UPBs in bipartite quantum system to construct a family UPBs in multipartite quantum systems.
\bigskip
\section{Preliminary}

\noindent{\textbf{Definition\cite{BDM+99,DMS+03}.}} A set of states
\{$|\phi_{i}\rangle\in\mathbb{C}^{m}\bigotimes\mathbb{C}^{n}:i=1,2,\cdots,k,\,k<mn$\}
is called a $k$-member UPB if and only if

(i) $|\phi_{i}\rangle$, $i=1,2,\cdots,k$, are product states;

(ii) $\langle\phi_{i}|\phi_{j}\rangle=\delta_{ij}$;

(iii) if $\langle\phi_{i}|\psi\rangle=0$ for all $i=1,2,\cdots,k$, then $|\psi\rangle$ cannot be product state.

 Now  we present the results of the minimal number of UPBs in bipartite quantum system\cite{CJ15,AL01}.

\begin{lemma}\label{min}
Let $f(d_1,d_2)$ be the smallest possible number of states in a UPB in $\mathbb{C}^{d_1}\otimes\mathbb{C}^{d_2}(d_1, d_2\geq 3)$. Then:\\
(1) If either $d_1$ or $d_2$ is odd, then $f(d_1,d_2)=d_1+d_2-1.$\\
(2) If both of $d_1$ and $d_2$ are even, then $f(d_1,d_2)=d_1+d_2.$
\end{lemma}

In addition, there is a family of UPBs for general bipartite quantum system \cite{DMS+03}.

\begin{lemma}\label{UPBmn}
In $\mathbb{C}^{m}\otimes\mathbb{C}^{n}$ with $n>3,m\geq3$ and $n\geq m$, there is a UPB of size $mn-2m+1$.
\end{lemma}

\section{UPBs in bipartite quantum system}

In this section, we mainly pay attention to the construction of UPB in bipartite quantum system. Firstly, we give a general method to construct a UPB from some old ones as follow. Suppose $\{|i\rangle \}_{i=0}^{m-1}$ and $\{|j\rangle \}_{ j=0}^{n-1}$ are orthogonal basis of quantum system $A$ and $B$ respectively. A pure state of the bipartite system is $|\psi\rangle=\sum m_{i,j}|ij\rangle$, then we call the $m\times n$ matrix $M_{|\psi\rangle}$ whose $(i,j)$ entry is $m_{(i-1),(j-1)}$  the corresponding matrix of $|\psi\rangle$. It is easy to show that $|\psi\rangle$ is a product state if and only if the rank of $M_{|\psi\rangle}$ is $1$.

\begin{theorem}\label{basic1}
 Suppose in $\mathbb{C}^m\otimes\mathbb{C}^{n_i}$, there is a UPB with $k_i$ missing states  number for $i=1,2$, then there is a UPB in $\mathbb{C}^m\otimes\mathbb{C}^{n_1+n_2}$ with $k_1+k_2$ missing states number. If we replace one of the UPB by a complete orthogonal product base and corresponding $k_i$ by $0$. The conclusion still holds.
\end{theorem}
\noindent \emph{Proof:} We only prove the first statement, the second can be proved similarly. Suppose that $\{|i\rangle_A\mid i=0,1,\ldots,m-1\}$ and $\{|j\rangle_B\mid j=0,1,\ldots,n_1+n_2-1\}$ are standard normal basis of system $A$ and $B$ respectively. Set
{
$$
\begin{array}{l}
  L=\text{span}_{\mathbb{C}}\{|ij\rangle\mid 0\leq i\leq m-1; 0\leq j \leq n_1-1\}, \\
  R=\text{span}_{\mathbb{C}}\{|ij\rangle\mid 0\leq i\leq m-1; n_1\leq j \leq  n_1+n_2-1\}.
\end{array}
$$}
Then $L$ and $R$ can be looked as the bipartite systems $\mathbb{C}^m\otimes\mathbb{C}^{n_1}$ and $\mathbb{C}^m\otimes\mathbb{C}^{n_2}.$ Hence, we can suppose $S_L=\{|v_i\rangle\in L\mid i=1,2,\ldots,mn_1-k_1\}$ and $S_R=\{|w_j\rangle\in R\mid j=1,2,\ldots,mn_2-k_2\}$ are the UPB in $\mathbb{C}^m\otimes\mathbb{C}^{n_1}$ and $\mathbb{C}^m\otimes\mathbb{C}^{n_2}$ respectively. We claim that $S_L\cup S_R$ is a  UPB in $\mathbb{C}^m\otimes\mathbb{C}^{n_1+n_2}$.

Clearly, the states in $S_L\cup S_R$ are orthogonal with each other. Suppose $H_{S_L}=\text{span}_{\mathbb{C}}(S_L), H_{S_R}=\text{span}_{\mathbb{C}}(S_R)$ and $H_{S_L}\oplus H_{S_L}^{\perp}=L, H_{S_R}\oplus H_{S_R}^{\perp}=R$. That is, $H_{S_L}^{\perp}$  (resp.$H_{S_R}^{\perp}$)is the set of states in $L$(resp.$R$) that is orthogonal all the states in $H_{S_L}$( resp.$H_{S_R} $). Then we have the following orthogonal decomposition:
$$V=L\oplus R=H_{S_L}\oplus H_{S_L}^{\perp}\oplus H_{S_R}\oplus H_{S_R}^{\perp}.$$
Since $S_L\cup S_R$ are orthogonal with each other, then we have $H_{S_L\cup S_R}=\text{span}_{\mathbb{C}}(S_L\cup S_R)=H_{S_L}\oplus H_{S_R}.$
By the above orthogonal decomposition , $H_{S_L\cup S_R}^{\perp}\supseteq H_{S_L}^{\perp}\oplus H_{S_R}^{\perp}.$ Since the states $S_L$ and $S_R$ are linearly independent, we can obtain that
$$\text{dim}_{\mathbb{C}} ( H_{S_L\cup S_R}^{\perp}) =  \text{dim}_{ \mathbb{C} }( H_{S_L}^{\perp} )+ \text{dim}_{ \mathbb{C} }( H_{S_R}^{\perp} ).$$
 With this equality and the above inclusion relation, we obtain $H_{S_L\cup S_R}^{\perp}= H_{S_L}^{\perp}\oplus H_{S_R}^{\perp}.$ Then we only need to show that there are no product states in $H_{S_L\cup S_R}^{\perp}$. Now for any nonzero vector $|\alpha\ra\in H_{S_L\cup S_R}^{\perp}, \ |\alpha\ra=a|v\ra+b|w\ra$ for $|v\ra\in H_{S_L}^{\perp}, |w\ra\in H_{S_R}^{\perp}.$
If we consider the corresponding matrix to $|\alpha\ra.$ Suppose $M_1$ is the corresponding matrix of $a|v\ra$,  $M_2$ is the corresponding matrix of $b|w\ra,$
then $M=(M_1|M_2)$ is the matrix corresponding to $|\alpha\ra.$ Since one of $a|v\ra$ or $b|w\ra$ is nonzero, which is equivalent to $\text{rank}(M_1)\geq 2$ or $\text{rank}(M_2)\geq2$. Since both two cases deduce that $\text{rank}(M)\geq 2,$ so $|\alpha\ra$ is not a product state. Hence $S_L\cup S_R$ is a  UPB in $\mathbb{C}^m\otimes\mathbb{C}^{n_1+n_2}$.\qed

\bigskip

\noindent{\bf Example 1.} Let $L=\text{span}_{\mathbb{C}}\{|ij\ra\mid i=0,1,2; j=0,1,2\}$, $R=\text{span}_{\mathbb{C}}\{|ij\ra\mid i=0,1,2; j=3,4,5\},$ then $$L\cong \mathbb{C}^3\otimes\mathbb{C}^3,\ \ R\cong \mathbb{C}^3\otimes\mathbb{C}^3.$$  Let $S_L=\{|0\ra|0-1\ra, |0-1\ra|2\ra, |1-2\ra|0\ra, |2\ra|1-2\ra, |0+1+2\ra|0+1+2\ra\}, \ S_R=\{|0\ra|3-4\ra, |0-1\ra|5\ra, |2\ra|4-5\ra, |1-2\ra|3\ra, |0+1+2\ra|3+4+5\ra\}.$ Then $S_L$ is a UPB in $L$, $S_R$ is a UPB in $R$. By the theorem \ref{basic1}. we have $S_L\cup S_R$ is a UPB in $\mathbb{C}^6\otimes\mathbb{C}^6$. The specific states present  in FIG.1.

\begin{figure}[h]\label{state}
\normalsize
\includegraphics[width=0.40\textwidth,height=0.30\textwidth]{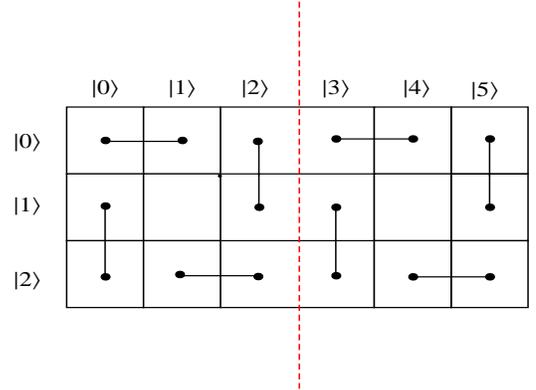}

 \caption{The construction of UPB in $\mathbb{C}^{6}\bigotimes\mathbb{C}^{6}$ by two subsystems $\mathbb{C}^{3}\bigotimes\mathbb{C}^{3}$. The left part representation the system $L$, the right part representation the system $R$. The state $|0+1+2\ra|0+1+2\ra$ and $|0+1+2\ra|3+4+5\ra$
  are not shown on this figure.}\label{Example}
\end{figure}
\begin{theorem}\label{basic2}
 If there is a  UPB in $\mathbb{C}^{m_i}\otimes\mathbb{C}^n$ with $k_i$ missing states  number  for $i=1,2$, then there is a UPB in $\mathbb{C}^{m_1+m_2}\otimes \mathbb{C}^n$ with $k_1+k_2$ missing states number.
  \end{theorem}

\noindent \emph{Proof:}  The proof is similar with the argument of theorem \ref{basic1}.\qed

\begin{corollary}\label{foursquare}
 If there is a UPB in $\mathbb{C}^{m_i}\otimes\mathbb{C}^{n_j}$ with $k_{ij}$ missing states number($i,j=1,2$), then there is a UPB in $\mathbb{C}^{m_1+m_2}\otimes\mathbb{C}^{n_1+n_2}$ with $k_{11}+k_{12}+k_{21}+k_{22}$ missing states number.
 \end{corollary}

Now we apply the above method to derive families of UPBs in bipartite quantum system.

\medskip

\begin{theorem}\label{nn}
 In $\mathbb{C}^{n}\otimes\mathbb{C}^m(7\leq n\leq m \leq n+1)$, there are UPBs with missing states number varying from $4$ to $\lfloor\frac{mn}{2}\rfloor+1$.
\end{theorem}

\noindent \emph{Proof:}  We prove by induction on $n$ such that in $\mathbb{C}^{n}\otimes\mathbb{C}^m$ ($n\leq m \leq n+1$) there are UPBs with missing states number varying from $4$ to $\lfloor\frac{mn}{2}\rfloor+1$.

We have checked the statement for $n=7,8,\ldots,13 $(see the table in APPENDIX). By induction the statement holds for $7\leq k\leq n-1$, we need to prove that it also holds for $n$. We can suppose $n\geq 14$.

Now we consider the case $\mathbb{C}^{n}\otimes\mathbb{C}^m $ with $n\leq m\leq n+1$.  There is a unique expression as the sum of two integers $n=n_1+n_2$ such that $n_1\leq n_2\leq n_1+1$.  Similarly, we write $m=m_1+m_2$ with similar condition.  It is not difficult to check that $|n_i-m_j|\leq 1$  and $n_i\geq7, m_j\geq7$ for all $i,j\in\{1,2\}$. Then $\mathbb{C}^{n }\otimes\mathbb{C}^{m}$ can be looked as
\begin{center}
\begin{tabular}{|c|c|}
  \hline
  $\mathbb{C}^{n_1}\otimes\mathbb{C}^{m_1}$ & $\mathbb{C}^{n_1}\otimes\mathbb{C}^{m_2}$ \\\hline
  $\mathbb{C}^{n_2}\otimes\mathbb{C}^{m_1}$ & $\mathbb{C}^{n_2}\otimes\mathbb{C}^{m_2}$ \\
  \hline
\end{tabular}
\end{center}

Then by induction, there are UPBs in $\mathbb{C}^{n_i}\otimes\mathbb{C}^{m_j} $ with missing state number varying from 4 to $\lfloor\frac{n_i m_j}{2} \rfloor+1$.
Then by corollary \ref{foursquare}, we can obtain some UPBs in $\mathbb{C}^{n}\otimes\mathbb{C}^{m} $  with missing state number varying from 16 to $\sum_{i,j=1}^2(\lfloor\frac{n_i m_j}{2} \rfloor+1).$ Clearly, there are also UPBs in $\mathbb{C}^{n}\otimes\mathbb{C}^{m} $  with missing states number varying from 4 to 16 which can be seem as induced from $\mathbb{C}^{7}\otimes\mathbb{C}^{7} $. Moreover, we have the following inequalities
$$
  \begin{array}{lcl}
    \displaystyle\sum_{i,j=1}^2(\lfloor\frac{n_i m_j}{2} \rfloor+1) & \geq & \displaystyle\sum_{i,j=1}^2( (\frac{n_i m_j}{2}-\frac{1}{2}) +1)\\
      & = &(\displaystyle\sum_{i,j=1}^2 \frac{n_i m_j}{2})+2 \\
      & \geq & \lfloor\frac{n  m }{2} \rfloor+1. \\
  \end{array}
$$
The first inequality holds for $\lfloor\frac{N}{2} \rfloor$ is equal to $\frac{N}{2}$ when $N$ is even and equal to $\frac{N-1}{2} $ when $N$ is odd.
 \qed

 \medskip

 The following figure   show another proof  of theorem \ref{nn}, we can use the UPBs  in the first triangle to obtain the UPBs in the second triangle. Then use the second one to obtain the third one and so on.

\begin{widetext}

\begin{figure}[h]
\normalsize
\includegraphics[width=0.97\textwidth,height=0.50\textwidth]{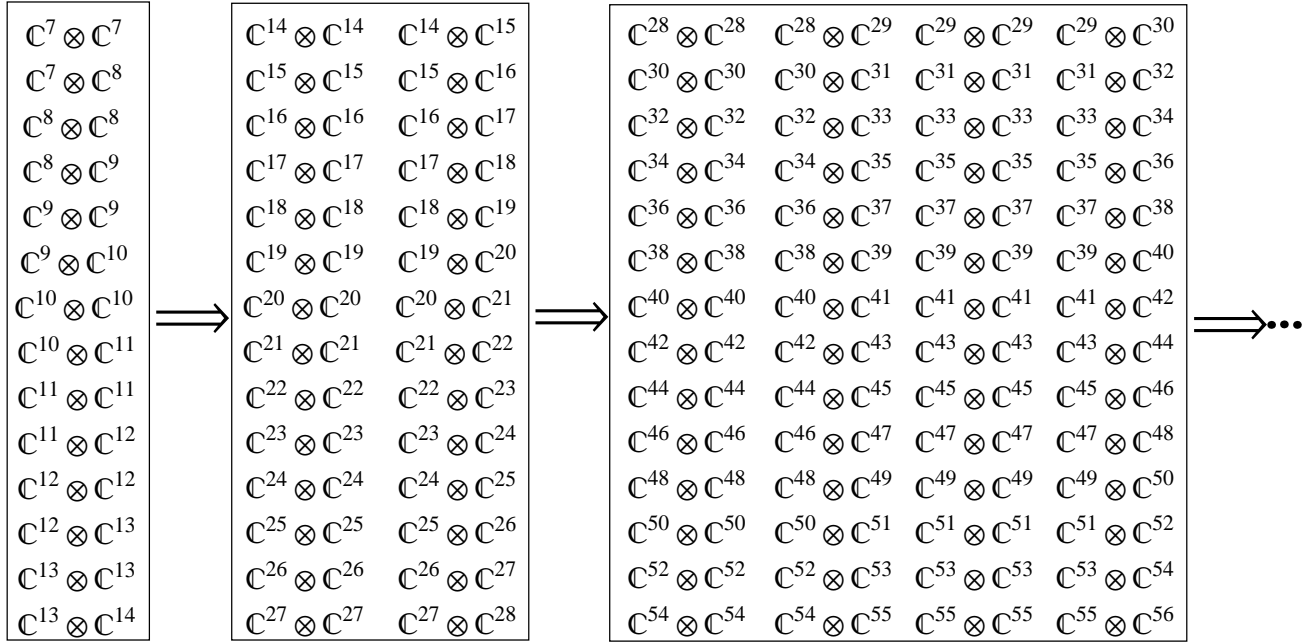}

 \caption{Another idea of the   proof  of theorem \ref{nn}.}\label{Example}
\end{figure}
\end{widetext}

\begin{theorem}\label{tensor7}
 In $\mathbb{C}^{n}\otimes\mathbb{C}^7(7\leq n)$, there are UPBs with missing states number $4,5,..., 6n-14,6n-12,6n-6$.
 \end{theorem}
 \noindent \emph{Proof:} Now we prove the above statement by induction. The cases n=7,8,9,10 have been calculated in the table of APPENDIX.  Suppose for any
 $n$ with $7\leq n\leq k$, the statement in the theorem is correct. We should use this to prove that it is also correct for the case $n=k+1$. Moreover, we can assume that $k\geq10$. Then $\mathbb{C}^{k+1}\otimes\mathbb{C}^7$ can be  seem as
  $$(\mathbb{C}^{3}\otimes\mathbb{C}^7)\oplus(\mathbb{C}^{k-2}\otimes\mathbb{C}^7) \text{ or } (\mathbb{C}^{4}\otimes\mathbb{C}^7)\oplus(\mathbb{C}^{k-3}\otimes\mathbb{C}^7).$$
   By induction, there are UPBs in $ \mathbb{C}^{k-2}\otimes\mathbb{C}^7$ (resp. $ \mathbb{C}^{k-3}\otimes\mathbb{C}^7$) with missing state number  $4,5,...,6k-26,6k-24,6k-18$(resp.$4,5,...,6k-32,6k-30,6k-24$). After considering the UPBs in $\mathbb{C}^{3}\otimes\mathbb{C}^7$ and $\mathbb{C}^{4}\otimes\mathbb{C}^7$ and by using theorem \ref{basic1}, we can obtain UPBs in $\mathbb{C}^{k+1}\otimes\mathbb{C}^7$ with missing number $ 4,5,...,6k-8,6k-6$. At last, the UPB with missing state number equal to $6k$ are derived from the minimal number UPB as showed in lemma \ref{min} .\qed

\begin{theorem}\label{odd}
 In $\mathbb{C}^{2m+1}\otimes\mathbb{C}^n(m\geq2,  n \geq10)$, there are UPBs with missing states number varying from $4$ to $2m(n-8)$.
 \end{theorem}

\noindent \emph{Proof:}
$\mathbb{C}^{2m+1}\otimes\mathbb{C}^n$ can be  seem as
   $(\mathbb{C}^{2m+1}\otimes\mathbb{C}^{n-7})\oplus(\mathbb{C}^{2m+1}\otimes\mathbb{C}^7)$.
We notice that, by lemma \ref{min}, there is a UPB in $\mathbb{C}^{2m+1}\otimes\mathbb{C}^k$ ($3\leq k\leq n-7$) with number $2m+1+k-1$, hence with missing states number $(2m+1)k-(2m+1)-k+1=2mk-2m=2m(k-1).$
Hence all these can extends to be UPBs in $\mathbb{C}^{2m+1}\otimes\mathbb{C}^{n-7}$ with missing states number $2m(k-1)\ (3\leq k\leq n-7)$.
Moreover, by theorem \ref{tensor7}, in $\mathbb{C}^{2m+1}\otimes\mathbb{C}^7$ hence in $\mathbb{C}^{2m+1}\otimes\mathbb{C}^n$ , there are UPBs with missing states  number varying from $4$ to $12m-8$,  where $12m-8> 4m+4$ when $m\geq 2.$

For any $4m+4 <l < 2m(n-8),$ $l$ can be expressed as $l=q (2m)+r$ with $0\leq r<2m,$ $2\leq q<n-8$. In the following, according to the condition of $r$, we express $l$  as a sum whose first term are reachable missing states number in $\mathbb{C}^{2m+1}\otimes\mathbb{C}^{n-7}$ and second term  are reachable missing states number in $\mathbb{C}^{2m+1}\otimes\mathbb{C}^{7}$. Then by theorem \ref{basic1}, we obtain a UPB in $\mathbb{C}^{2m+1}\otimes\mathbb{C}^n$ with $l$ missing state number.

(i) If $r=0$, then  $l=q (2m)+0$.

(ii)If $0<r<4$, then $l=(q-1) (2m)+(2m+r).$

(iii)If $4\leq r<2m$, then $l=q (2m)+r$.
\qed

\begin{theorem}\label{even}
 In $\mathbb{C}^{2m}\otimes\mathbb{C}^n(m\geq2,  n \geq10)$, there are UPBs with missing number varying from $4$ to $(2m-1)(n-8)$.
 \end{theorem}

\noindent \emph{Proof:} The proof is just similar with theorem 3.  In $\mathbb{C}^{2m}\otimes\mathbb{C}^k$ there are UPB with $2m+k$ states, hence with $2mk-2m-k=2m(k-1)-(k-1)-1=(2m-1)(k-1)-1$ missing states.
Hence all these can extend to UPB in $\mathbb{C}^{2m}\otimes\mathbb{C}^{n-7}$ with $(2m-1)(k-1)-1$ ($3\leq k \leq n-7$) missing states number.
Moreover, by theorem \ref{tensor7}, there are UPBs in $\mathbb{C}^{2m}\otimes\mathbb{C}^7$ with missing states number varying from $4$ to $12m-14$,  where $12m-14\geq 4m+1$ when $m\geq 2.$

For any $4m+1 <l < (2m-1)(n-8)-1,$ $l$ can be expressed as $l+1=q (2m-1)+r$ with $0\leq r<2m-1$ and $2\leq q<n-8.$ In the following, according to the condition of $r$, we express $l$  as a sum whose first term are reachable missing states number in $\mathbb{C}^{2m }\otimes\mathbb{C}^{n-7}$ and second term  are reachable missing states number in $\mathbb{C}^{2m }\otimes\mathbb{C}^{7}$. Then by theorem \ref{basic1}, we obtain a UPB in $\mathbb{C}^{2m }\otimes\mathbb{C}^n$ with $l$ missing states number.

(i)If $r=0$, then $l=(q (2m-1)-1)+0$.

(ii) If $0< r<4$,   $l=((q-1) (2m-1)-1)+(2m-1+r).$

(iii)If $4\leq r<2m-1$, then $l=(q (2m-1)-1)+r$.
\qed

\begin{corollary}\label{bipartite}
 In $\mathbb{C}^{m}\otimes\mathbb{C}^n$ $(m\geq4,  n \geq10)$,  there are UPBs with missing states number varying from $4$ to $(m-1)(n-8)$.
 \end{corollary}
 \noindent \emph{Proof:}Combining the conclusion of theorem \ref{odd} and theorem \ref{even}, we get the conclusion.\qed

\section{UPBs in multipartite quantum systems}

In this section, we first give a method which is originated from ref.\cite{J14} to construct a UPB of $N$-partite quantum systems by some UPBs in the ($N-1$)-partite quantum systems. Then combining this method with the UPBs we have obtained in bipartite quantum system above, we give a family of UPB in multipartite quantum system.

\begin{lemma}\label{mul_lemma}
 Suppose there are $K(K\geq2)$ sets $\{\mathcal{P}^{(i)}\}$ of UPBs in $\mathbb{C}^{d_1}\otimes\mathbb{C}^{d_2}\otimes...\otimes\mathbb{C}^{d_N}$ and we denote them as
$\mathcal{P}^{(i)}=\{|\psi_j^{(i)}\rangle,\ j=1,2,...,n^{(i)}\}$. Then the set $\mathcal{P} =\{|\psi_j^{(i)}\rangle|i-1\rangle,i=1,2,...,K,\ j=1,2,...,n^{(i)}\}$ is a UPB in
$\mathbb{C}^{d_1}\otimes\mathbb{C}^{d_2}\otimes...\otimes\mathbb{C}^{d_N}\otimes\mathbb{C}^{K}$.
\end{lemma}

\noindent \emph{Proof:} Clearly, the states in $\mathcal{P}$ are all product states of $\mathbb{C}^{d_1}\otimes\mathbb{C}^{d_2}\otimes...\otimes\mathbb{C}^{d_N}\otimes\mathbb{C}^{K}$. Moreover, they are orthogonal. So we only need to prove that there are no more product states  in $\mathcal{P}^\perp$. Suppose $|\phi_1\rangle|\phi_2\rangle...|\phi_N\rangle|\phi_{N+1}\rangle$ is a nozero product state in $\mathcal{P}^\perp$. Since $|\phi_{N+1}\rangle$ is nonzero, there is an $i_0$  with $1\leq i_0\leq K$ such that $\langle i_0-1|\phi_{N+1}\rangle\neq0$. Then
the orthogonality between $|\phi_1\rangle|\phi_2\rangle...|\phi_N\rangle|\phi_{N+1}\rangle$ and $|\psi_j^{(i_0)}\rangle|i-1\rangle , j=1,2,...,n^{(i_0)}$ give
  $$  (\langle\psi_j^{(i_0)}|)(|\phi_1\rangle|\phi_2\rangle...|\phi_N\rangle)\langle i_0-1|\phi_{N+1}\rangle=0.  $$
  These imply that $|\phi_1\rangle|\phi_2\rangle...|\phi_N\rangle$ is an product state in $\mathbb{C}^{d_1}\otimes\mathbb{C}^{d_2}\otimes...\otimes\mathbb{C}^{d_N}$ that is orthogonal to all the states in $\mathcal{P}^{(i_0)}$. However, this is contradicted with our assumption that $\mathcal{P}^{(i_0)}$ is a UPB. In conclusion, $\mathcal{P}$ is a UPB. \qed

  \medskip

After replacing some sets $\{\mathcal{P}^{(i)}\}$  by the full product basis with at least a set is still a UPB, then $\{\mathcal{P} \}$ is also a UPB.

\begin{theorem}\label{mult}
If in $\mathbb{C}^{d_1}\otimes\mathbb{C}^{d_2}\otimes...\otimes\mathbb{C}^{d_N} $ $(d\geq2)$ there are UPBs with missing states number from $4$ to $L$ with  $L\geq8$, then there are UPBs in $\mathbb{C}^{d_1}\otimes\mathbb{C}^{d_2}\otimes\mathbb{C}^{d_3}\otimes\ldots\otimes\mathbb{C}^{d_{N+1}}$ with missing states number varying from $4$ to $d_{N+1}L$.
\end{theorem}
\noindent \emph{Proof:}  Given any integer $m$ with $4\leq m\leq d_{N+1}L$, there is unique expression $m=qL+r, 0\leq r<L$. By lemma \ref{mul_lemma}, in order to construct a UPB in $\mathbb{C}^{d_1}\otimes\mathbb{C}^{d_2}\otimes\mathbb{C}^{d_3}\otimes\ldots\otimes\mathbb{C}^{d_{N+1}}$ with missing states number $m$, we only need to construct $ d_{N+1}$ sets of UPBs with missing states number $\{m_i\}_{i=1}^{d_{N+1}}$ in $\mathbb{C}^{d_1}\otimes\mathbb{C}^{d_2}\otimes\mathbb{C}^{d_3}\otimes\ldots\otimes\mathbb{C}^{d_{N }}$ such that  $m=\sum_{i=1}^{d_{N+1}}m_i.$  In the following, we separate our the construction of the $ d_{N+1}$ sets of UPBs in $\mathbb{C}^{d_1}\otimes\mathbb{C}^{d_2}\otimes\mathbb{C}^{d_3}\otimes\ldots\otimes\mathbb{C}^{d_{N }}$ into three cases according to the number $r$.

(i) $4\leq r<L$. In this case, $0\leq q <d_{N+1}$. Let $\{\mathcal{P}^{(i)}\}_{i=1}^q$ be  $q$ sets of  UPBs   whose missing states number are all $L$. Moreover, let $\mathcal{P}^{(q+1)}$ be a UPB with missing states number $r$ and for any integer $q+1<i<d_{N+1}$ we let $\mathcal{P}^{(i)}$ be a set of full orthogonal product basis.

(ii) $0<r<4$. In this case, $1\leq q<d_{N+1}$. Then $m=(q-1)L+(L-(4-r))+4$. Let $\{\mathcal{P}^{(i)}\}_{i=1}^{q-1}$ be  $q-1$ sets of  UPBs  whose missing states number are all $L$. Moreover, let $\mathcal{P}^{(q )}$ be a UPB with missing states number $L-(4-r)$ whose existence is guaranteed  by the assumption and the condition $4\leq L-(4-r)\leq L$ and $\mathcal{P}^{(q+1 )}$ be a UPB with missing states number $4$. And for any integer $q+1<i<d_{N+1}$ we let $\mathcal{P}^{(i)}$ be a set of full orthogonal product basis.

(iii)$r=0$. That is, $m=qL$. This is the easiest case for it can derive from $q$ sets of UPBs with $L$ missing states number and $d_{N+1}-q$ sets of full orthogonal product basis in $\mathbb{C}^{d_1}\otimes\mathbb{C}^{d_2}\otimes...\otimes\mathbb{C}^{d_N}$. \qed

\medskip

 Now we use the above theorem \ref{mult} and lemma \ref{mul_lemma} to give lots of UPBs in multipartite quantum systems.

 \begin{corollary}
  Suppose that  $d_1\geq d_2\geq d_3\geq...\geq d_N$ and $d_1\geq 10, d_2\geq 4, N\geq 3$ . Then in $\mathbb{C}^{d_1}\otimes\mathbb{C}^{d_2}\otimes...\otimes\mathbb{C}^{d_N} $,  there are UPBs with missing states number  varying from $4$ to $(d_1-8)(d_2-1)\prod_{i=3}^Nd_i$.
  \end{corollary}

\section{conclusion and discussion}
We give a method to obtain   new UPBs from the old ones. In order to show the power of the method, we use it to construct UPBs with different kinds of number.
Firstly, we show that in $\mathbb{C}^{n}\bigotimes\mathbb{C}^{n}$ there are   UPBs with missing states number varying from $4$ to $\frac{n^2}{2}$. Then we  study the  more general quantum system $\mathbb{C}^{m}\otimes\mathbb{C}^n(m\geq4,  n \geq10)$. And we obtain that there are UPBs with missing states number varying from $4$ to $(m-1)(n-8).$ Moreover, use this results we can also obtain some results of the multipartite quantum systems.

We notice that there is some gap between the minimal number $f(m,n)$  of UPB and the number $mn-(m-1)(n-8)$. Hence it is very interesting to investigate whether there are some other UPBs whose numbers can full fill this gap. Moreover, it is well know that the UPBs can be used to construct bounded entangled states\cite{BDM+99}. Hence using our results, one can obtain some bounded entangled states with different ranks. Hence it is valuable to derive some other methods that can full fill this gap which enable us to construct bounded entangled states with ranks varying from $4$ to $mn-f(m,n)$.

\bigskip

$Acknowledgments:$ This work is supported by the NSFC 11475178, NSFC 11571119 and NSFC 11275131.

\clearpage

\section{appendix}
{
\small
\begin{tabular}{|l|l|l|l|l|l|l|l|l|l|l|l|l|}
\multicolumn{12}{c}{\tabincell{c}{\textbf{Table 1:} By using the UPBs in lemma \ref{min}, lemma \ref{UPBmn} and those in $\mathbb{C}^3\otimes\mathbb{C}^4 $ in ref.\cite{YGX+15},  we calculate\\
   the missing state number of UPBs in $\mathbb{C}^m\otimes\mathbb{C}^n $ for small $m,n$ by our method.}}\\\hline

   & 3 & 4 & 5 & 6 & 7 & 8 & 9 & 10 & 11 & 12 & 13 & 14 \\\hline
  3 & 4 &  &  &  &  & &  &  &  &  & & \\\hline
  4 & 4-6 & 4-8  &  &  &  & &  &  &  &  &  & \\\hline
  5 &4-6, 8 &4-8, 12  &\tabincell{c}{4-9, 12,\\ 16}  &  &  & &  &  &  &  & &  \\\hline
  6 & 4-6, 8, 10 &4-12, 14   &\tabincell{c}{4-14, 16,\\ 20}  &\tabincell{c}{4-16, 18,\\20, 24}  &  & &  &  &  &  & &  \\\hline
  7 &  \tabincell{c}{4-6, 8-10,\\ 12 }&4-14, 18   &\tabincell{c}{4-18, 20,\\ 24}  &\tabincell{c}{4-22, 24,\\ 30}  &\tabincell{c}{4-28, 30,\\ 36}  & &  &  &  &  &  & \\\hline
  8 & \tabincell{c}{4-6, 8-12,\\ 14} &4-18, 20   &\tabincell{c}{4-22, 24,\\ 28}  &\tabincell{c}{4-26, 28,\\ 30, 34}  &\tabincell{c}{4-34, 36,\\ 42}  &\tabincell{c}{4-40, 42,\\48} &  &  &  &  &  & \\\hline
  9 & \tabincell{c}{4-6, 8-14,\\16} &4-20, 24   &\tabincell{c}{4-26, 28,\\ 32}  &\tabincell{c}{4-32, 34,\\40}  &\tabincell{c}{4-40, 42,\\ 48}  & \tabincell{c}{4-46, 48,\\ 56 } &\tabincell{c}{4-54, 56,\\ 64}  &  &  &  &  & \\\hline
  10 & \tabincell{c}{4-6, 8-16,\\ 18} &4-24, 26   &\tabincell{c}{4-30, 32,\\36}  & \tabincell{c}{4-36, 38,\\40, 44}  & \tabincell{c}{4-46, 48,\\ 54}  &\tabincell{c}{4-54, 56,\\62} &\tabincell{c}{4-62, 64,\\72}  &\tabincell{c}{4-70, 72,\\80}  &  &  &  & \\\hline
  11 & \tabincell{c}{4-6, 8-18,\\20} &4-26, 30   &\tabincell{c}{4-34, 36,\\ 40}  &\tabincell{c}{4-42, 44,\\50}  &\tabincell{c}{4-52, 54,\\60}  &\tabincell{c}{4-60, 62,\\ 70} &\tabincell{c}{4-70, 72,\\ 80}  &\tabincell{c}{4-78, 80,\\ 90}  &\tabincell{c}{4-88,\\90, 100}  &  & &  \\\hline
  12 & \tabincell{c}{4-6, 8-20,\\22} &4-30, 32   &\tabincell{c}{4-38, 40,\\ 44}  &\tabincell{c}{4-46, 48,\\50, 54}  &\tabincell{c}{4-58, 60,\\ 66}  &\tabincell{c}{4-68, 70,\\ 76} &\tabincell{c}{4-78, 80,\\ 88}  &\tabincell{c}{4-88, 90,\\ 98}  &\tabincell{c}{4-98,\\100, 110}  &\tabincell{c}{4-108,\\110, 120}  &  & \\\hline
  13 & \tabincell{c}{4-6, 8-22,\\  24} &4-32, 36   &\tabincell{c}{4-42, 44,\\ 48}  &\tabincell{c}{4-52, 54,\\ 60}  &\tabincell{c}{4-64, 66,\\ 72}  &\tabincell{c}{4-74, 76,\\ 84} &\tabincell{c}{4-86, 88,\\ 96}  &\tabincell{c}{4-96, 98,\\ 108}  &\tabincell{c}{4-108,\\110, 120}  &\tabincell{c}{4-118,\\120, 132}  &\tabincell{c}{4-130,\\132, 144} & \\\hline
  14 &\tabincell{c}{4-6, 8-24,\\ 26} &4-36, 38   &\tabincell{c}{4-46, 48,\\ 52}  &\tabincell{c}{4-56, 58,\\60, 64 } &\tabincell{c}{4-70, 72,\\ 78}  &\tabincell{c}{4-82, 84,\\ 90} &\tabincell{c}{4-94, 96,\\ 104}  &\tabincell{c}{4-106,\\108, 116}  &\tabincell{c}{4-118,\\120, 130}  &\tabincell{c}{4-130,\\132, 142}  &\tabincell{c}{4-142,\\144, 156} & \tabincell{c}{4-154,\\156, 168} \\\hline
\end{tabular}

}
\medskip

\end{document}